\documentclass[twocolumn,showpacs,preprintnumbers,amsmath,amssymb]{revtex4}

\usepackage{graphicx}
\usepackage{bm}

\begin{document}

\title{Local Leaders in Random Networks}


\author{V.D. Blondel$^{1}$\email{blondel@inma.ucl.ac.be} }

\author{J.-L. Guillaume$^{1}$\email{guillaume@inma.ucl.ac.be} }

\author{J.M. Hendrickx$^{1}$\email{julien.hendrickx@uclouvain.be} }

\author{C. de Kerchove$^{1}$\email{dekerch@inma.ucl.ac.be} }

\author{R. Lambiotte$^{2}$\email{Renaud.Lambiotte@ulg.ac.be} }

\affiliation{
$^1$ INMA, Universit\'e catholique de Louvain,  4 avenue Georges Lemaitre,
B-1348 Louvain-la-Neuve, Belgium \\
$^2$ GRAPES, Universit\'e de Li\`ege, Sart-Tilman, B-4000 Li\`ege, Belgium }

\begin{abstract}
We consider local leaders in random uncorrelated networks, i.e. nodes whose degree is higher or equal than the degree of all of their neighbors. An analytical expression is found for the probability of a node of degree $k$ to be a local leader.
This quantity is shown to exhibit a transition from a situation where high degree nodes are local leaders to a situation where they are not when the tail of the degree distribution behaves like the power-law $\sim k^{-\gamma_c}$ with $\gamma_c=3$.
Theoretical results are verified by computer simulations and the importance of finite-size effects is discussed.
\end{abstract}


\maketitle

\section{Introduction}

In the last few years, the study of networks has received increasing attention from the scientific community~\cite{review1,review2} in disciplines as diverse as biology (metabolic and protein interactions), computer and information sciences (the Internet and the World Wide Web), etc. It has been shown that many empirical networks differ from regular lattices by their random structure and by the heterogeneity of the node properties, i.e. nodes may exhibit very different topological properties inside the same network. The best known case is node degree heterogeneity which results in fat-tailed degree distributions where many nodes are sparsely connected while a few nodes, or hubs, receive a large number of links~\cite{bara}. It is now well-known that degree heterogeneity~\cite{boguna,sood} and, especially the presence of hubs, are important factors that may radically alter the propagation of {\em data}, e.g.  rumours~\cite{viral}, opinions~\cite{galam,lambiotte} or viruses~\cite{virus} and may provoke its weakness in front of targeted attacks~\cite{pastor0,may}.

\begin{figure}
\includegraphics[width=2.9in]{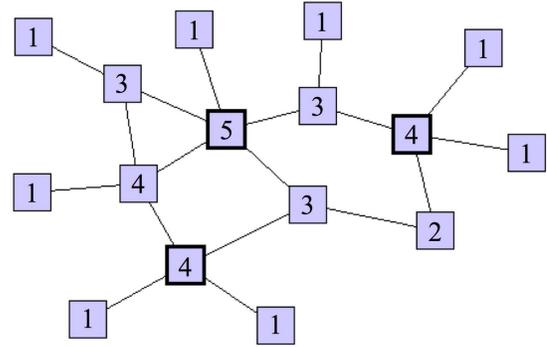}
\vspace{-0.cm}
\caption{Sketch of a random network composed of 16 nodes. The network possesses 3 local leaders, two of them being strict leaders.}
\label{fig1}
\end{figure}

The important role played by hubs in the above processes has therefore motivated a detailed study of the extremal properties of networks. Different contributions~\cite{mda,kr} have focused on the properties of the degree of the leader, i.e. the node with the highest degree, on the probability that the leader never changes and on related leadership statistics~\cite{tl}. These approaches, based on the theory of extreme statistics~\cite{jg}, have provided an excellent description of the behaviour of the global extrema in the network but, surprisingly, the statistics of local extrema have not been considered yet.  There are several reasons, though, to focus on {\em local leaders}, namely nodes whose degree is larger or equal to the degree of their neighbors and on {\em strict leaders}, namely nodes whose degree is strictly larger than the degree of their neighbors (see Fig.~\ref{fig1}). Such nodes may be viewed as local hubs that trigger the communication between nodes at the local level. Indeed, individuals usually compare their {\em state} (e.g. opinion, wealth, idea, etc.) with the {\em state} of their direct neighbors, thereby suggesting that a local leader might have a preponderant role in its own neighborhood, whatever the absolute value of its connectivity. As a rich among the poor, a local leader might therefore have a more dominant role than as a rich among the richest. From a marketing point of view, for instance, the identification of such nodes might be of interest in order to target nodes that play an important role within {\em circles of friends}~\cite{social}. Let us also stress that local leaders form a subset of nodes that might grasp important characteristics of the whole network and could be helpful in order to visualize its internal features.

In this paper, we focus on the properties of local leaders in uncorrelated random networks, i.e. networks where the degrees of neighboring nodes are not correlated~\cite{assor}. In section II, we derive an analytical formula for the probability $P_k$ for a node of degree $k$ to be a local leader and show that this probability undergoes a phase transition where the control parameter is the degree distribution itself~\cite{control}. When the tail of the distribution decreases faster than a power-law $\sim k^{-\gamma_c}$ with $\gamma_c=3$, the probability to be a local leader goes to 1 for large enough values of $k$. When the tail of the distribution decreases slower than $\sim k^{-\gamma_c}$, in contrast, this probability vanishes for large enough degrees. In section III, we verify our theoretical predictions by computer simulations and show how finite size effects may affect the above transition. In section IV, finally, we conclude and propose generalizations of the concept of local leader.

\section{Being rich among the poor, and vice versa}

Let us consider an undirected random network determined by its degree distribution $n_k$, i.e. the probability that a randomly chosen node has degree $k$. By construction, this distribution satisfies the relations
\begin{eqnarray}\label{eq:defbase}
\sum_{k=1}^{\infty} n_k =1, ~~ \sum_{k=1}^{\infty} k n_k = z,
\end{eqnarray}
where $z=2 L/N$ is the average degree, $N$ the total number of nodes and $L$ the total number of links in the network.
In the above relations, we have assumed that there are no nodes with degree $k=0$, which is reasonable as such nodes are excluded from the network structure.

Let us now evaluate the probability $P_k$ that a node of degree $k$ is a local
leader - the case of strict leaders will be briefly discussed at
the end of this section. To do so, one first has to look at the
probability $q_j$ that a neighbor of the node under consideration
has a degree $j$. In a network where the degrees of adjacent nodes
are statistically independent, it is well-known that $q_j$ is
equal to the probability that a randomly chosen link arrives at a
node of degree $j$, so that 
$q_j =  j n_j/z.$
The probability for this node to have a degree $j \leq k$ is
therefore
\begin{eqnarray}
\label{uni} q^{'}_k = \frac{\sum_{j=1}^{k} j n_j}{z}.
\end{eqnarray}
By definition, a node with degree $k$ is a local leader if all of
its $k$ neighbors have a degree smaller or equal to $k$. By using
the statistical independence of the degrees of these $k$
neighbors, $P_k$ is found by multiplying (\ref{uni}) $k$ times
\begin{eqnarray}
\label{expression} P_k = \left( \frac{ \sum_{j=1}^{k} j
n_j}{z}\right)^k.
\end{eqnarray}

In general, $P_k$ is a function of $k$ whose behaviour may be
evaluated numerically by inserting the degree distribution $n_k$
of the network in Eq.(\ref{expression}) and by performing the
summations. In the following, however, we would like to derive
general properties of $P_k$ that do not depend on the details of
$n_k$. To do so, let us only focus on the asymptotic behaviour of
$P_k$, when $k$ is large, and assume that $n_k$ may be
approximated for large enough values of $k$ by a power-law: 
$n_k = C k^{-\gamma}$, where $C$ is a normalization constant. The
case of pure power-laws where $n_k = C k^{-\gamma}$ for all $k$ will 
be detailled later on.


Let us emphasize that such a tail of the degree distribution is very general, as it includes scale-free distributions ($\gamma$ finite), while exponential distributions are recovered in the limit $\gamma \rightarrow \infty$. In the following, we focus on general values of $\gamma$, with the sole constraint that  $\gamma > 2$ so that the average degree is well-defined. In that case, $\sum_{j=1}^{\infty} j n_j=z$ is a finite number and Eq.(\ref{expression}) may be rewritten as
\begin{eqnarray}
\label{expression2}
P_k = \left( 1 - \frac{ \sum_{j=k+1}^{\infty} C j^{-(\gamma-1)}}{z}\right)^k,
\end{eqnarray}
where we used the fact that
$
 \sum_{j=1}^{k} j n_j = \sum_{j=1}^{\infty} j n_j - \sum_{j=k+1}^{\infty} j n_j
$.

For large enough values of $k$, the summation in (\ref{expression2}) may be replaced by an integral so that $P_k$ asymptotically behaves like
\begin{eqnarray}
\label{expression3}
P_k \approx \left( 1 - \frac{ C k^{-(\gamma-2)}}{(\gamma-2) z}\right)^k.
\end{eqnarray}
In order to determine the asymptotic behaviour of $P_k$, it is useful to rewrite Eq.(\ref{expression3}) as
\begin{eqnarray}
\label{expression4}
P_k \approx e^{k \ln \left( 1 -  \frac{C k^{-(\gamma-2)}}{(\gamma-2) z}\right)}
\end{eqnarray}
whose dominating term is, when $k^{-(\gamma-2)}$ is sufficiently
small,
\begin{eqnarray}
\label{expression4bis}
P_k \approx e^{ \frac{-C k^{-(\gamma-3)}}{(\gamma-2) z}}.
\end{eqnarray}
By construction, $\gamma>2$ and $z$ is positive, so that the asymptotic values of $P_k$, for large enough values of $k$, is
\begin{equation}
\label{transition}
P_k \rightarrow
\begin{cases}
1      & {\rm for} \quad \gamma > 3,\cr
e^{ -  C/z}          & {\rm for} \quad \gamma = 3,\cr
 0 & {\rm for} \quad \gamma < 3,
\end{cases}
\end{equation}
The system therefore undergoes a transition at $\gamma=3$.
If the tail of the degree distribution decreases fast enough, so that 
$\gamma>3$, the probability $P_k$ asymptotically goes to~1. Consequently,
nodes with a higher degree have a larger probability to be local leaders.
When $\gamma<3$, in contrast, the probability to be a local leader decreases 
with the degree~$k$ and asymptotically vanishes, so that, surprisingly, 
nodes with a larger degree might have a smaller probability to be local 
leaders.

This result, which may appear intriguing at first sight, can be
explained by analyzing the competition between two trends. On one
hand, a node with a high degree has a higher probability of
having a higher degree than any other particular node, which
tends to increase its probability of being a degree leader. On the
other hand, a node with a higher degree has more neighbors,
which tends to decrease the probability of having a higher degree
than all its neighbors (see the exponent~$k$ in Eq.~(\ref{expression})).
Depending on the value of~$\gamma$, the asymptotic behaviour is
dictated by the first or the second phenomenon, with a transition
when $\gamma=3$ where an equilibrium occurs between the two
phenomena.

One should also note that the above calculations simplify in term of Harmonic functions
$H(k,\gamma) \equiv \sum_{i=1}^{k} i^{-\gamma}$,
when the degree distribution is a pure power-law $n_k=C k^{-\gamma}$ for all $k$, where $C=1/\sum_{k=1}^{\infty}k^{-\gamma}=1/H(\infty,\gamma)$.  Indeed, in that case, the probability to be a local leader $P_k$ reads
\begin{eqnarray}
\label{expressionH}
P_k = \left( \frac{ \sum_{j=1}^{k}  j^{-(\gamma-1)}}{\sum_{j=1}^{\infty}  j^{-(\gamma-1)}}\right)^k = \left(\frac{H(k,\gamma-1)}{H(\infty,\gamma-1)}\right)^k.
\end{eqnarray}
Using the asymptotics of the harmonic numbers~\cite{knuth}
\begin{eqnarray}
\label{expressionHbis}
H(k,\gamma-1) = H(\infty,\gamma-1) - \frac{ k^{-(\gamma-2)}}{(\gamma-2)}
\end{eqnarray}
valid when $\gamma>2$, it is straightforward to recover the
transition~(\ref{transition}) where $e^{ -  C / z}$ is now given by
$e^{ -  1/ H(\infty,2)}=e^{ - 6/ \pi^2}$, since $z=H(\infty,\gamma-1)/H(\infty,\gamma)$.

Before going further, let us discuss the case of strict leaders. In that case, 
the calculations are the same as previously, except for the sums in~$P_k$ that 
do not go until~$k$ but until~$k-1$. However, this difference is vanishingly 
small for large enough values of~$k$, so that the transition~(\ref{transition}) 
is recovered.

\section{Simulations and finite size effects}

In this section, we verify the validity of the theoretical predictions (\ref{expression}) and, especially, the existence of the regime $P_k \rightarrow 0$ when $\gamma<3$. One should first stress that the results derived in the previous section are valid for uncorrelated networks composed of an infinite number of nodes. However, whatever the specified degree distribution $n_k$, a typical realization of the network (in a computer simulation or in realistic situation)  involves only a finite number of nodes. This also implies that the largest degree $k_{max}$ in the network is a finite number. The degree $k_{max}$ of this global leader might be estimated by using tools from the theory of extreme statistics~\cite{jg}, but the main point here is that the global leader is also a local leader. Consequently, the probability for a node of degree $k_{max}$ to be a local leader, when measured in such a system, is $P_{k_{max}}=1$, in contradiction with the prediction $P_k \rightarrow 0$.

\begin{figure}[!h]
\includegraphics[width=3.2in]{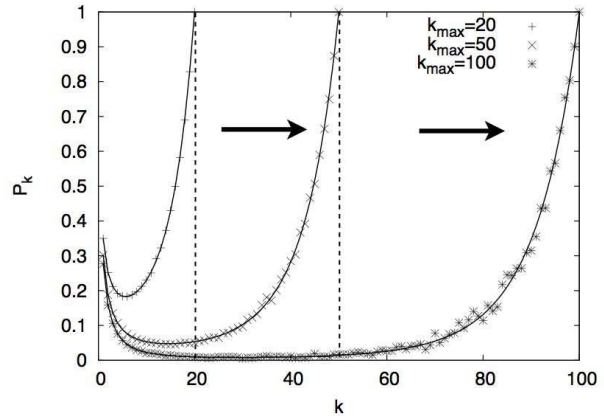}
\caption{$P_k$ measured in random networks composed of $10^5$ nodes and whose degree distribution is a truncated power law (\ref{truncated}) with $\gamma=2.2$. The results are averaged $100$ times. The solid lines are the theoretical prediction (\ref{expression}), evaluated numerically for the degree distributions (\ref{truncated}). The value of $k$ where $P_k$ begins to increase toward $P_k=1$ due to finite-size effects (see main text) is seen to be proportional with $k_{max}$.}
\label{fig2}
\end{figure}

In order to highlight this finite-size effect with computer simulations, it is helpful to consider the truncated power laws defined by
\begin{eqnarray}
\label{truncated}
n_k &=& D k^{-\gamma}  ~~{\rm for} \quad k \leq k_{max},\cr
n_k &=& 0  ~~{\rm otherwise},
\end{eqnarray}
where the constant of normalization depends on $\gamma$ and on the cut-off $k_{max}$, $D=1/\sum_{k=1}^{k_{max}} k^{-\gamma}$. Such degree distributions offer the possibility to tune the value of the extremal degree $k_{max}$ together with a particularly simple expression for $n_k$.
To generate numerically random uncorrelated networks with the specified degree distribution (\ref{truncated}), we proceed as follows~\cite{finite}.  We assign to each node $i$ in a set of $N$ nodes a degree $k_i$ sampled from the probability distribution (\ref{truncated}) and impose that $\sum_{i=1}^N k_i$ is even. Then, the network is constructed by randomly assigning the $L=\sum_{i=1}^N k_i/2$ edges while respecting the pre-assigned degrees $k_i$. In the simulations, we have considered networks with $N=10^5$ nodes and averaged the results over 100 realizations of the random process. One should also stress that we have only considered truncated distributions such that $k_{max}$ is effectively the maximum degree for each realization of the network, i.e. such that the expected number of nodes with $k_{max}$ verifies $N n_{k_{max}}\geq 1$. Computer simulations (see Fig.~\ref{fig2}) show an excellent agreement with the theoretical prediction (\ref{expression}) and confirm that $P_k$ first decreases to values close to 0 when $\gamma<3$, as predicted by (\ref{transition}), before increasing to 1 due to finite size effects.

In order to evaluate where finite size effects become non-negligible, we have focused on the value $k_{c}$ where $P_{k}$ is minimum and studied the relation between $k_c$ and $k_{max}$. By inserting the distribution (\ref{truncated}) and integrating numerically (\ref{expression}), one observes that $k_c$ increases linearly with $k_{max}$, $k_{c} \approx  \alpha k_{max}$. When $\gamma=2.2$, for instance, one finds $\alpha=0.3189$. This linear dependence has important consequences as it implies that finite size effects only affect a vanishingly small number of the nodes when $k_{max}$ is sufficiently large. To show so, let us consider the proportion $n_{FS}$ of nodes affected by the finite size effects

\begin{eqnarray}
\label{affected}
n_{FS} &=& \sum_{k=\alpha k_{max}}^{k_{max}} D k^{-\gamma} \approx \int_{k= \alpha k_{max}}^{k_{max}} D k^{-\gamma} \cr
&=& \frac{D}{\gamma-1} (\alpha^{-(\gamma-1)} - 1) k_{max}^{-(\gamma-1)},
\end{eqnarray}
where the summation has been replaced by an integral, as $k_{max}$ is sufficiently large.
The quantity $n_{FS}$ obviously goes to zero when $k_{max} \rightarrow \infty$.

Before concluding, let us also derive the behaviour of $P_k$ close to $k_{max}$. In that case, numerical integration shows an exponential decrease in $(k_{max}-k)$ so that one looks for a solution of the form
\begin{eqnarray}
\label{zkap}
P_{k} \approx e^{E (k_{max}-k)},
\end{eqnarray}
 where the constant $E$ is a found by comparing (\ref{zkap}) with

\begin{eqnarray}
\label{zkap2}
P_k = e^{ k \ln\left(1 - \frac{\sum_{j=k+1}^{k_{max}} D j^{-(\gamma-1)}}{z} \right)},
\end{eqnarray}
and by looking at the dominant terms for small values of $k^{'} \equiv k_{max}-k$. When $k_{max}$ is sufficiently large, it is straightforward to show that

 \begin{eqnarray}
\label{zkap3}
E &\approx& k_{max} \ln(1) + k_{max-1} \ln(1 - D k_{max}^{-(\gamma-1)} /z)\cr
&\approx& - D k_{max}^{-(\gamma-2)}/z.
\end{eqnarray}
 This asymptotic behaviour has been successfully compared with computer simulations.

\section{Conclusion}

In this paper, we have analysed the statistical properties of local leaders. Such nodes, that may be viewed as local hubs, have a crucial location in a social or information network, as they dominate all their neighbors. Their identification and a better understanding of their properties might therefore be of practical interest. In marketing, for instance, local leaders are good candidates to target in order to maximize a marketing campaign or to minimize the erosion of customers from a company, e.g. {\em churn} for mobile operators~\cite{churn}. We have observed that the probability for a node of degree $k$ to be a local leader undergoes a transition from a {\em rich is rich} to a {\em rich is poor} situation, that suggests that nodes with a high degree might not be the most influential at the local level. It is interesting to stress that the transition takes place at a realistic value of the power-law exponent $\gamma_c=3$~\cite{GNC,lambi}, i.e. scale-free distributions usually have an exponent between $2$ and $3$~\cite{ne}, and that  $\gamma_c=3$ is also the critical value under which
the variance diverges.
To conclude, one should stress that the local maxima of other node quantities could also give insight into the network structure, e.g. the number of triangles~\cite{social}. More general definitions of local leaders could also be considered, e.g. a node of degree $k$ is a $\alpha$-leader if all of its neighbors have a degree $k^{'}<k/\alpha$.  A generalization of our study to such situations and a comparison with empirical data (where nodes might exhibit degree-degree correlations) could therefore be of interest.

\medskip
\noindent
{\bf Acknowledgements}


V. Blondel, J.-L. Guillaume, J. Hendrickx
and C. de Kerchove are supported by the Concerted Research Action (ARC)
''Large Graphs and Networks'' from the ''Direction de la recherche
scientifique - Communaut\'e fran\c{c}aise de Belgique.'', by the
EU HYCON Network of Excellence (contract number FP6-IST-511368),
and by the Belgian Programme on Interuniversity Attraction Poles
initiated by the Belgian Federal Science Policy Office. The
scientific responsibility rests with its authors. Julien Hendrickx
is FNRS fellow (Belgian Fund for
Scientific Research) R. Lambiotte has been supported by European
Commission Project CREEN FP6-2003-NEST-Path-012864.\\

The research described in this paper was initiated during a
research retreat organized by Rapha\"el Jungers and Vincent
Blondel in Matagne la Petite, Belgium in 2007.

\end{document}